\documentclass[twocolumn,aps,pre]{revtex4}

\usepackage{dcolumn}
\usepackage{bm}
\usepackage[dvips]{graphicx}

\newcommand{\xv}{{\mathbf x}}
\newcommand{\qv}{{\mathbf q}}
\newcommand{\pv}{{\mathbf p}}
\newcommand{\nv}{{\mathbf n}}
\newcommand{\Nv}{{\mathbf N}}
\newcommand{\Ha}{{\mathcal H}}

\newcommand{\gradv}{{\bm{\nabla}}}
\newcommand{\pop}{{\hat{\pi}}}

\begin{document}
\title{The Structure of TGB$_C$ Phases}
\author{Arindam Kundagrami and T.C. Lubensky}
\affiliation{Department of Physics, University of Pennsylvania,
Philadelphia, Pennsylvania 19104}
\date{\today}
\begin{abstract}
We study the transition from the cholesteric phase to two TGB$_C$
phases near the upper critical twist $k_{c2}$: the Renn-Lubensky
TGB$_C$ phase, with layer normal rotating in a plane perpendicular
to the pitch axis, and the Bordeaux TGB$_C$ phase, with the layer
normal rotating on a cone parallel to the pitch axis. We calculate
properties, including order-parameter profiles, of both phases.
\end{abstract}
\pacs{PACS:61.30.-v, 61.30.Dk, 64.70.Md} \maketitle

\begin{figure}
\centerline{\includegraphics{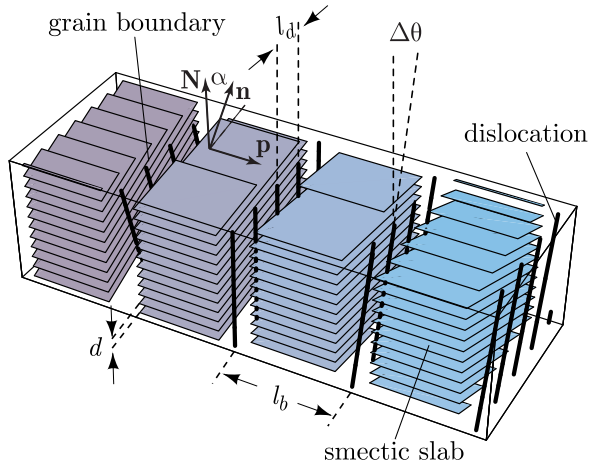}}
\caption{\label{fig1}The RL TGB$_C$ phase.
There is a fixed angle between the layer normal $\Nv$ and the
director $\nv$, which rotate in the plane perpendicular to the
pitch axis $\pv$.}
\end{figure}

\begin{figure}
\centerline{\includegraphics{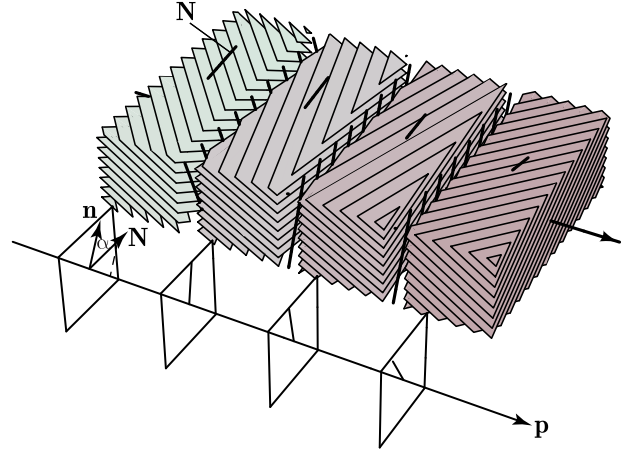}}
\caption{\label{fig2}The Bordeaux TGB$_C$ phase.
There is a fixed angle between $\Nv$ and $\nv$, but $\nv$ rotates
in the plane perpendicular to $\pv$, and $\Nv$ rotates on a cone
whose axis is parallel to $\pv$ }
\end{figure}

Smectic liquid crystalline phases\cite{deGennes} are layered
structures: they are fluid-like in two-dimensions and solid-like
in the third. Twist-grain-boundary or TGB
phases\cite{RennLub89,Pindak89,review} are phases of smectic
liquid crystals induced by molecular chirality.  They consist of
periodically spaced grain boundaries, each composed of an array of
parallel dislocations, separating smectic slabs as depicted in
Figs.\ \ref{fig1} and \ref{fig2}. The layer normals $\Nv$ of the
slabs rotate in discrete jumps across the grain boundaries. These
remarkable phases are the analog in liquid
crystals\cite{deGennesSc} of the Abirkosov flux lattice in
superconductors\cite{SC} with the complex smectic
mass-density-wave amplitude $\psi$ the analog of the
superconducting order parameter, dislocations in the grain
boundary the analog of vortices, and the chiral coupling constant
$h$ induced by molecular chirality the analog of the external
magnetic field $H$.

The simplest TGB phase is the TGB$_A$ phase in which the smectic
slabs between grain boundaries have the character of a bulk
smectic-$A$ phase in which the layer normal $\Nv$ and the director
$\nv$, specifying the direction of average molecular alignment,
are parallel to each other in a plane perpendicular to the pitch
axis along $\pv$.   In TGB$_C$ phases, the smectic slabs have the
character of a bulk smectic-$C$ with $\nv$ tilted relative to
$\Nv$.  Two distinct structures for the TGB$_C$ phase immediately
come to mind.  In the first\cite{RennLub-C,RennC}, both $\Nv$ and
$\nv$  rotate in the plane perpendicular to $\pv$ but with a
finite angle between them as shown in Fig.\ \ref{fig1}. We will
refer to this as the Renn-Lubensky or RL TGB$_C$ phase. In the
second TGB$_C$ phase, first discovered\cite{BTGBC} and
subsequently studied in detail by the Bordeaux group  and
collaborators\cite{BandC}, $\nv$ rotates in the plane
perpendicular to $\pv$, but $\Nv$ rotates on a cone with a
component parallel to $\pv$ so that $\pv$ does not lie parallel to
the smectic layers as shown in Fig.\ \ref{fig2}. We will refer to
this as the Bordeaux or B TGB$_C$ phase. No pure form of the RL
phase has been reported, though phases with two-dimensional
modulation of the local RL TGB$_C$ structure have been
observed\cite{squareTGBC}.  Though the RL TGB$_C$ structure may be
unstable with respect to these modulations, we assume here that it
can be stable. We will not discuss the TGB$_{C^*}$
phase\cite{RennC} with smectic-$C^*$ slabs in which the director
rotates in a cone from layer to layer.

Our goal is to study the structure of both the RL and B phases
near the upper critical field $h_{c2}$ where the TGB$_C$ phase
becomes unstable with respect to the cholesteric phase in which
the smectic order parameter is zero and the director twists in a
helical fashion about $\pv$ with a pitch $P$. We follow closely
the procedure developed by Abrikosov\cite{SC} in his analysis of
the superconducting flux phase near the upper critical field
$H_{c2}$ and applied successfully to the TGB$_A$ phase near
$h_{c2}$\cite{RennLub89}. Our analysis of the transition to the
bordeaux TGB phase is essentially identical to that presented by
Luk'yanchuk \cite{Luk98}. Using a more general model than his,
which does not permit a stable RL phase, we study both the B and
RL phases, including their order-parameter profiles, but not the
TGB$_{2q}$ phase he introduced.

\begin{figure}
\centerline{\includegraphics{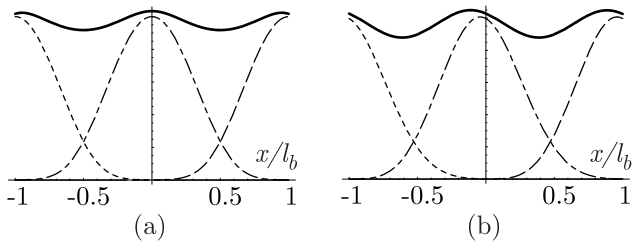}}
\caption{\label{fig3} The square amplitude $|\psi(x,0,0)|^2$smectic
order parameter as a function of $x$ in (a) the Bordeaux and (b)
the RL TGB$_C$ phases. These figures show the squared slab
wavefunctions $|\phi(x-n l_b)|^2$ with $n=-1$ centered at $-l_b$
(short dash, short space) with $n=1$ centered at $l_b$ (long dash,
short space) and with $n=0$ (short and long dashes) and
$|\psi(x,0,0)|^2$ (full line).  Though $|\phi(x)|$ dies off fairly
rapidly, $|\psi(x,0,0)|^2$ has a robust value at grain-boundary
positions $x/l_b = \pm 0.5$.  Note the asymmetry about $x=0$ in
the RL case.}
\end{figure}

Several results of our analysis are worthy of note.  The linear
stability operator associated with the TGB$_A$ phase, like that
associated with the Abrikosov phase, is a quantum harmonic
oscillator Hamiltonian, $\Ha_u = -d^2/du^2 + u^2$, where $u$ is a
rescaled coordinate along $\pv$. The same operator associated with
the B phase is a $u^4$-anharmonic oscillator Hamiltonian,
$-d^2/du^2 + u^4$, whereas that associated with the RL phase is
the dual of the Bordeaux operator, $d^4/du^4 + u^2$. The
dependence of the grain-boundary spacing $l_b$, the dislocation
spacing $l_d$ within a grain boundary, and the smectic coherence
length $\xi$ on the cholesteric pitch $P$ and the layer spacing
$d$ is different in the three phases as reviewed in Table
\ref{table1}. The near equality of $l_b$ and $l_d$ in the TGB$_A$
phase and their $P^{1/2}d^{1/2}$ dependence on pitch and layer
spacing has been verified\cite{Pindak89}.
Experimentally\cite{BandC}, $l_b$ is substantially larger than
$l_d$ in the Bordeaux TGB$_C$ phase in agreement with Table
\ref{table1}. Finally, we find that the smectic order parameter,
though depressed at the grain boundaries, is reasonably constant
in the two TGB$_C$ phases as shown in Fig.\ \ref{fig3}. We find no
evidence in the B phase of melted grain boundaries along which
$\psi \sim 0$ as suggested by Dozov\cite{Dozov}.

\begin{table}
\begin{tabular}{|l|c|c|c|c|} \hline
Phase& $\Ha_u$& $l_b$ & $l_d$ & $\xi$  \\ \hline

TGB$_A$&$-d^2/du^2 +u^2$& $P^{1/2}d^{1/2}$ & $P^{1/2}d^{1/2}$
& $P^{1/2}d^{1/2}$ \\
Bordeaux& $-d^2/du^2 + u^4$  & $P^{2/3} d^{1/3}$ & $P^{1/3}
d^{2/3}$ & $P^{2/3} d^{1/3}$\\
RL & $d^4/du^4 + u^2$& $P^{1/3} d^{2/3}$ & $P^{2/3} d^{1/3}$
&$P^{2/3} d^{1/3}$ \\
\hline
\end{tabular}
\caption{$\Ha_u$ and proportionality of $l_b$, $l_d$,
and $\xi$ to powers of $P$ and $d$ in TGB phases.}
\label{table1}
\end{table}

To describe the smectic properties of the TGB phases near
$h_{c2}$, we use a slight modification of the Chen-Lubensky (CL)
model\cite{Chen-Lub} for the nematic-to-smectic-$A$ and
nematic-to-smectic-$C$ transitions.  In this model, the smectic
molecular number density is expressed as $\rho = \psi + \psi^*$,
where $\psi$ is the complex mass-density-wave amplitude with
wavenumbers with magnitude peaked near $q_0= 2 \pi/d$.  The free
energy, $F_{\psi} = F_H + F_{NL}$ is the sum of a nonlinear part,
$F_{NL} = \frac{1}{2} g\int d^3 x |\psi|^4$, and a part harmonic
in $\psi$,
\begin{eqnarray}
F_H &= &\int d^3 x [{\tilde r} |\psi|^2 + D_{||}
|[\nabla_{||}^2(\xv)
+
q_0^2] \psi|^2  \nonumber \\
& & + D_{\perp}|[\nabla_{\perp}^2(\xv) + q_{0\perp}^2]\psi|^2\\
& & +D_{||\perp} [\nabla_{||}^2(\xv) + q_0^2]\psi^*
[\nabla_{\perp}^2(\xv) + q_{0\perp}^2]\psi + {\rm c.c.}
],\nonumber
\label{Fpsi}
\end{eqnarray}
where $\nabla_{||}^2 ( \xv) \equiv (\nv( \xv) \cdot \gradv )^2$
and $\nabla_{\perp}^2 ( \xv ) = \nabla^2 - \nabla_{||}^2 ( \xv )$.
With the identification $q_{0\perp}^2 = -C_{\perp}/2D_{\perp}$ and
${\tilde r} = r - C_{\perp}^2/(4 D_{\perp})$, this model is simply
a alternative representation of the CL model\cite{Chen-Lub} except
for the term proportional to $D_{\|| \perp}$, which, as we shall
see, is needed to stabilize the RL phase.  When $q_{0 \perp}^2<0$,
this model has a phase transition from the nematic to the
smectic-$A$ phase with $\psi = \psi_A e^{i q_0 z}$, where $\psi_A
= (-r/g)^{1/2}$ . When $q_{0\perp}^2 >0$, it has transition to the
Sm$C$ phase with $\psi = \psi_C e^{i \qv_C \cdot \xv}$, where
$\psi_C = (-\tilde{r} /g)^{1/2}$, with $\qv_C = (q_{0\perp}
\cos\gamma,q_{0\perp} \sin\gamma, q_0)$ for any angle $\gamma$.

To complete the description of our system, we add the Frank free
energy, $F_{\nv}$, for the director including the contribution
from molecular chirality:
\begin{eqnarray}
F_{\nv} & = & \frac{1}{2} \int d^3 x \{K_1 (\gradv \cdot \nv )^2 +
K_2 [\nv \cdot( \gradv \times \nv )]^2 \nonumber \\
& & + K_3 [\nv \times( \gradv \times \nv )]^2 + h \nv \cdot(
\gradv \times \nv ) \}.
\label{fn}
\end{eqnarray}
When $\psi = 0$, the equilibrium state is the cholesteric phase
determined by $F_{\nv}$ alone with director
\begin{equation}
\nv_0 ( x ) = ( 0, - \sin k_0 x , \cos k_0 x ) ,
\label{nv0}
\end{equation}
where $k_0 = h/K_2 \equiv 2 \pi /P$.

There are several dimensionless quantities in $F = F_{\psi} +
F_{\nv}$ that play a role in our analysis. One is the ratio
$k_0/q_0 = d/P$, which is of order $10^{-2}$ or less. Our primary
concern will be the limit in which $k_0/q_0$ approaches zero, and
we will consider only leading terms in this ratio. Other
parameters are the ratios,
\begin{displaymath}
\eta_{\perp} = D_{\perp}/D_{||} , \qquad \eta_{\||\perp} =
D_{||\perp}/D_{||} , \qquad \omega = q_{\perp 0}^2 /q_0^2
\end{displaymath}
and the twist Ginzburg parameter, $\kappa_2 = (g K_2/2)^{1/2}/(4
D_{||} q_0^3)$. In at least one material\cite{MITsmc},
$\eta_{||\perp} \approx 0$ and $\eta_{\perp} \ll 1$, but there is
no a priori reason why either of these conditions should not be
violated. $\omega = \tan^2 \alpha$ is a measure of the equilibrium
tilt angle $\alpha$ between $\nv$ and $\Nv$. It is more convenient
to use the twist $k_0 = h/K_2$ rather than $h$ as a measure of
chirality. The critical twist at which the cholesteric phase
becomes unstable to the TGB phases is $k_{c2} = h_{c2}/K_2$.

To determine when the cholesteric phase first becomes unstable
with respect to the development of smectic order and to find our
variational wave functions\cite{RennLub89} for the TGB phases, we
calculate the lowest eigenvalues and associated eigenfunctions of
the harmonic kernel obtained from $F_H$ with $\nv(\xv)$ replaced
by $\nv_0 ( x)$. This kernel $K$ is a periodic function of $x$
with period $P/2$. Its eigenfunctions are, therefore, plane waves
in the $yz$ plane that can be expressed as $\psi ( \xv ) =
\Phi_{\qv_P} ( x ) e^{i \qv_P \cdot \xv}$ where $\qv_P = (0, q_y,
q_z)$ and where, as indicated, the form of the function
$\Phi_{\qv_P} ( x )$ can depend on $\qv_P$. When $k_0=0$, the
eigenfunctions associated with the lowest eigenvalue of $K$ are
$\psi ( \xv ) = e^{i \qv_C \cdot \xv}$ in which $\qv_C$ can have a
nonvanishing $x$-component.  We allow explicitly for this
component of $\psi$ that varies periodically with $x$ by setting
$\Phi_{\qv_P} ( x )=\phi_{\qv_P}(x) e^{iq_x x}$ and $\psi ( \xv )
= \phi_{\qv_P} ( \xv ) e^{i \qv \cdot \xv}$ where $\qv = ( q_x,
\qv_P)$. With this form for $\psi$, $F_H$ becomes
\begin{equation}
F_H^0 = A\int dx \phi_{\qv_P}^* ( x ) \Ha ( x, \pop, \qv )
\phi_{\qv_P} ( x )
\label{F0phi}
\end{equation}
where $ \pop = i^{-1} d/dx$ is the momentum operator and
\begin{eqnarray}
\lefteqn{\Ha(x,\pop,\qv) =  {\tilde r}+ D_{||}Q_{||}^2 ( x, \qv) +
D_{\perp} Q_{\perp}^2 ( x, \pop,\qv )}\\
& & + D_{|| \perp}[ Q_{||}(x,\qv) Q_{\perp}( x, \pop,\qv )+
Q_{\perp} ( x, \pop,\qv ) Q_{||}(x,\qv) ] , \nonumber
\label{Hphi}
\end{eqnarray}
where $Q_{||} ( x,\qv) = q_{||}^2 ( x ) - q_0^2$ , with
$q_{||}^2(x) = (\qv \cdot \nv_0 ( x ) )^2$, and $Q_{\perp}( x,
\pop,\qv)= \pop^2+ 2 q_x \pop\ + q_{\perp}^2 (x) - q_{0\perp}^2$,
where $q_{\perp}^2 (x)= q^2 -q_{||}^2(x)$.

$\Ha(x,\pop,\qv)$ is a periodic function of $x$ with a band
spectrum and Bloch eigenfunctions. To lowest order in $k_0/q_0$,
however, eigenfunctions are localized at spatial minima in
${\tilde r}(x, \qv)=\Ha(x,0,\qv)$, which can be approximated by
the lowest-order terms in a Taylor expansion about these minima.
For any given $\qv$, ${\tilde r}(x, \qv)$ will have a minimum at
some $ x=x_m ( \qv)$. Since ${\tilde r}(x, \qv)$ depends on $x$
only in the combination $\nv_0 ( x) \cdot \qv$, $x_m (
U(\theta)\qv )= x_m (\qv) + \theta /k_0$, where $U(\theta)$ is the
operator that rotates $\qv_m$ through an angle $\theta$ about
$\pv$, it is always possible to find a $\qv = \qv_m$ such that
$x_m ( \qv_m ) = 0$.  If $\phi(x)$ is an eigenfunction of $\Ha (
x,\pop,\qv_m)$ with energy $\epsilon$, then $\psi ( \xv ) = \phi (
x - \theta/k_0) e^{i \qv_m ( \theta)\cdot \xv}$, where $\qv_m (
\theta ) = U(\theta) \qv_m$, is an eigenfunction of the harmonic
kernel of $F_H$ with energy $\epsilon$ for all $\theta$.

Our approach, therefore, is to find those $\qv$'s that minimize
${\tilde r} ( 0, \qv)$ or, equivalently, those $\qv$'s for which
$Q_{||}( 0 ) = $ and $Q_{\perp} ( 0, 0) = 0$.  Since $q_{||}^2 (
0) = q_z^2$, and $q_{\perp}^2(0) = q_x^2 + q_y^2$, it follows that
${\tilde r} ( 0, \qv )$ is at its minimum equal to $\tilde r$ for
$q_z= q_0$ and $(q_x, q_y) = q_{0\perp} (\cos \gamma, \sin
\gamma)$ for any $\gamma$.  The Bordeaux phase corresponds to
$\gamma = 0$ and the RL phase to $\gamma = \pi/2$.  Having found
$\qv_m$, we can expand $Q_{||}$ and $Q_{\perp}$ in powers of $x$
and $\pop$:
\begin{eqnarray}
Q_{||} & = & - 2 k_0 q_0 q_{0\perp} \sin \gamma x + (q_0^2 -
q_{0\perp}^2 \sin^2 \gamma) + ... \nonumber\\
Q_{\perp} & = & \pop^2 + 2 q_{0\perp} \cos \gamma \pop - Q_{||}
\label{Q-Q}
\end{eqnarray}
These expressions simplify in the Bordeaux and RL cases to
\begin{eqnarray}
Q_{||}^B &= &q_0^2 k_0^2 x^2, \qquad Q_{\perp}^B = 2 q_{0\perp}
\pop - q_0^2 k_0^2 x^2  \\
Q_{||}^{RL}& = &- 2 k_0 q_0 q_{0\perp} x , \qquad Q_{\perp}^{RL}=
\pop^2 - 2 k_0 q_0 q_{0\perp} x
\label{QB-QRL}
\end{eqnarray}
plus terms, which we show shortly, that yield corrections to the
lowest order terms in $(k_0/q_0)$.  If is clear from these
expressions that the B and RL phases enjoy a sort of duality
obtained by interchanging $x$ and $\pop$.  The Hamiltonian for the
B phase will have terms proportional to $x^4$, $\pop^2$ and $\pop
x^2 + x^2 \pop$, whereas that for the RL phase will have terms
proportional to $\pop^4$, $x^2$, and $x \pop^2 + \pop^2 x$.  In
the B case, the $\pop x^2 + x^2 \pop$ term can be removed by
transforming the wave function via $\phi_B(x) = \exp(i \mu_{B}
x^3) {\tilde \phi}_B ( x )$ for an appropriate choice of $\mu_B$
while the $x^2 \pop + \pop x^2$ term in the RL case can be removed
by transforming the Fourier transform $\phi_{RL} (k) = \int dx
e^{-i kx} \phi_{RL} (x)$ to $\exp( i \mu_{RL} k^3) {\tilde
\phi}_{RL}(k)$.  In both cases, the eigenfunction ${\tilde
\phi}_B(x)$ and ${\tilde \phi}_{RL} ( x) $ are localized near
$x=0$ over some characteristic length $l$, and it is convenient to
express them as functions of the unitless variable $u=x/l$.  This
leads to the Hamiltonians for the RL and B phases expressed to
lowest order in $u$ and $\pop_u= i^{-1} d/du$:
\begin{eqnarray}
\Ha_{RL}^0 - \tilde{r}& = & 4 D_{||} q_0^4 s_2 \omega (k_0 l)^2
\left[ u^2 + \frac{1}{(k_0 l)^6}\frac{k_0^4 \eta_{\perp} s_1}{4
q_0^4 s_2^2 \omega}
\pop_u^4 \right] , \nonumber\\
\Ha_{B}^0 -\tilde{r}& = &D_{||} q_0^4 s_1 (k_0 l)^4 \left[ u^4 +
\frac{1}{(k_0 l)^6} \frac{4 k_0^2 \eta_{\perp} \omega}{q_0^2 s_1}
\pop_u^2 \right] ,
\label{H0B}
\end{eqnarray}
where $s_1 = 1 - (\eta_{||\perp}^2/\eta_{\perp})$ and $s_2 = 1 +
\eta_{\perp} - 2 \eta_{||\perp}$. We can choose $l$ to make the
respective coefficients of $\pop_u^4$ and $\pop_u^2$ in
$\Ha_{RL}^0$ and $\Ha_{B}^0$ be unity:
\begin{eqnarray}
(k_0 l_{RL})^6 &= & (k_0/q_0)^4 \eta_{\perp} s_1/(4 s_2^2 \omega) , \nonumber\\
(k_0  l_B)^6 \,\,& = & 4 (k_0/q_0)^2 (\eta_{\perp} \omega/s_1) .
\end{eqnarray}
With these choices, $\Ha_{RL}^0 = \tilde{r} + E_0 (\eta s_2)^{1/3}
(k_0/q_0)^{4/3}[u^2 + \pop_u^4]$ and $\Ha_B^0 = \tilde{r} +
\eta_{\perp}^{2/3} E_0 (k_0/q_0)^{4/3}[ u^4 +\pop_u^2]$, where
$E_0= D_{||}q_0^4 [16 \omega^2 s_1]^{1/3}$, are duals to each
other with $u^2 +\pop_u^4$ and $u^4 + \pop_u^2$ having the same
lowest eigenvalue $\epsilon_0$.

The eigenvalues of both $\Ha_{RL}^0 - \tilde{r}$  and $\Ha_B^0 -
\tilde{r}$ scale as $\overline{D} q_0^4 (k_0 /q_0)^{4/3}$. Higher
order tems in $k_0 x$ and $\pop$ neglected in Eq.\ (\ref{Q-Q})
yield corrections to the dominant $(k_0/q_0)^{4/3}$ behavior of
both $\Ha_{B}^0 - \tilde{r}$ and $\Ha_{RL}^0 - \tilde{r}$ of order
$(k_0/q_0)^2$ or higher.  In addition $\Ha^0 - {\tilde r}$ scales
as $k_0/q_0$ when $\gamma \neq 0, \pi/2$, and corrections to
$\Ha_{B}^0 - \tilde{r}$ and $\Ha_{RL}^0 - \tilde{r}$ scale,
respectively, as $\gamma^2 (k_0/q_0)^{2/3}$ and $(\pi/2 -
\gamma)^2 (k_0/q_0)^{2/3}$.  Thus the B and RL phases always have
lower energy than phases with intermediate values of $\gamma$.

The cholesteric phase becomes unstable at $k_0 = k_{c2}$ when the
smallest eigenvalue of $\Ha^0$ becomes zero.  Thus
\begin{displaymath}
k_{c2}^{RL} =  \frac{q_0}{(\eta_{\perp}s_2)^{1/4}}
\left(\frac{|\tilde{r}|}{\epsilon_0 E_0}\right)^{3/4} , \;\;
k_{c2}^B = \frac{q_0}{\eta_{\perp}^{1/2}}
\left(\frac{|\tilde{r}|}{\epsilon_0 E_0}\right)^{3/4} ,
\end{displaymath}
and near $|\tilde{r}| = 0$, both $k_{c2}(RL)$ and $k_{c2}(B)$
scale as $|\tilde{r}|^{3/4}$.  Their ratio is
$k_{c2}^{B}/k_{c2}^{RL}= (s_2\eta_{\perp})^{1/4}$.  Thus,
$k_{c2}^B>k_{c2}^{RL}$ and the cholestreric phase becomes unstable
to the B phase before the RL phase when $s_2 = 1 + \eta_{\perp} -
2 \eta_{||\perp} > \eta_{\perp}$, i.e., when $\eta_{||\perp}
<1/2$, and it becomes unstable to the RL phase before the B phase
when $\eta_{||\perp} > 1/2$.  This means that the RL phase is not
stable in the original CL model in which $D_{||\perp} = 0$.
Stability of the nematic phase in the absence of chirality
requires $D_{||} D_{\perp} - D_{||\perp}^2 = D_{||}D_{\perp}s_1
>0$ or $\eta_{||\perp}^2 < \eta_{\perp}$.  It is clearly possible
to satisfy both this condition and $\eta_{||\perp} >1/2$, so that
there is a range of parameters for which the RL phase is stable.

The B eigenfunctions are of the form $\phi (u ) = e^{i \mu'_{B}
u^3} {\tilde \phi}_B (u)$ where  $(u^4
+ \pop_u^2) {\tilde \phi}_B ( u ) = \epsilon_0 {{\tilde\phi}_B(u)}$. The
RL eigenfunctions are of the form $\phi_{RL} (u) = \int(dk/2 \pi)
e^{-i \mu_{RL}' k^3} {\tilde \phi}_B ( k)$, where ${\tilde \phi}_B
(u)$ and ${\tilde \phi}_B(k)$ are identical functions of different
arguments. ${\tilde \phi}_B(u)$ can be obtained numerically using
the shooting method, and it, along with ${\phi_{RL}}(u)$, is
plotted in Fig.\ {\ref{fig4}.  The resulting eigenvalue is
$\epsilon_0 = 1.060357 ...$.  An excellent approximation to
$\phi_B(u)$ over the entire range of $u$ is ${\tilde \phi}(u) =
\exp [ - \frac{1}{2}A u^2 \sqrt{1 + \frac{4}{9}(u/A)^2}]$, where
$A = 1.035$. This function satisfies the requirement that ${\tilde
\phi}(u) \rightarrow \exp(- \frac{1}{3}|u^3|)$ as $|u|\rightarrow
\infty$.

\begin{figure}
\centerline{\includegraphics{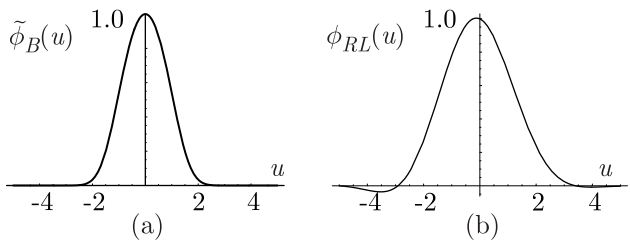}}
\caption{\label{fig4}Amplitude of eigenfunctions for
(a) the Bordeaux and (b) the RL case. Note that the RL
wavefunction has an oscillatory component and it is slightly
asymmetric.}
\end{figure}

To determine the structure of the TGB$_C$ phases, we construct
variational smectic order parameters from the degenerate set of
lowest eigenfunctions of $\Ha(x,\pop, \qv)$:
\begin{equation}
\psi ( \xv ) = C \sum_s \phi ( (x - n l_b ) /l) e^{i q_m ( \theta)
\cdot \xv} ,
\end{equation}
where $\qv_m ( \theta ) = U( \theta ) \qv_m $ and, as discussed in
the paragraph preceding Eq.\ (\ref{H0B}), $\qv_m$ has different
forms in the Bordeaux and RL phases. Following the treatment of
the cholesteric-to-TGB$_A$ transition\cite{RennLub89}, we can
write the total free energy after minimizing over director
fluctuations as
\begin{equation}
\frac{F}{K_2 q_0^2 V} = - \frac{1}{2} - \frac{\cal{A}}{4 \beta (
l_b / l)} \left(\frac{k_0 - k_{c2}}{k_0}\right)^2 ,
\end{equation}
where $\beta (l_b /l)$ depends on the separation $l_b$ between
grain boundaries, and $\cal{A}$ depends on $k_0$ and $l$ but not
on $l_b$.  Thus, the equilibrium value of $l_b$ is that which
minimizes $\beta(l_b/l)$, which can be expressed as $([\psi^4
]_{\text{av}}- \kappa_2^{-2}(k_0/q_0)^{2/3}
[f]_{\text{av}})/[\psi^2]_{\text{av}}^2$, where $f$ is a
complicated function of order $\psi^4$ and $[g]_{\text{av}}=
V^{-1} \int d^3 x g$ is the spatial average of $g$.  Carrying out
this minimization procedure using the analytic approximation for
$\phi_B(u)$, we find $l_b/l \approx 2.15$ for  the Bordeaux and
$l_b/l \approx 3.0$ for the RL TGB$_C$ phases, respectively. To
find the spacing between dislocations in a grain boundary, we use
the geometric relation $ k_0 = {d}/({\sin \alpha_0  l_b l_d})$,
where $\pi/2 - \alpha_0$ is the angle between $\Nv$ and $\pv$. Our
results for $l_b \sim l$, $l_d \sim P d/l$ and $\xi \sim (D q_0^2/
\tilde{r})^{1/2} \sim d^{1/3} P^{2/3}$ are summarized in table
\ref{table1}.  The wave function $\psi$ for our calculated values
of $l_b$ for both TGB$_C$ phases are shown in Fig.\ \ref{fig3}.

We have presented an overview of the properties of the the
Bordeaux and RL TGB$_C$ phases and the transition to them from the
cholesteric phase obtained from an Abrikosov-like analysis near
the upper critical twist $k_{c2}$ at which the cholesteric phase
becomes unstable. In a future publication\cite{KL2}, we will
present more details of our calculations, including a discussion
of the transition from type I to type II behavior. We will also
discuss the relation between our work and that of
Dozov\cite{Dozov}.

We thank Randall Kamien for a careful reading of this manuscript.
This work was supported in part by the National Science Foundation
under grant DMR 00-96531.


\begin{thebibliography}{10}
\bibitem{deGennes} P.G. de Gennes and J. Prost, {\it The Physics of
Liquid Crystals}, Second Edition (Clarendon Press, Oxford, 1993).
\bibitem{RennLub89} S. R. Renn, T. C. Lubensky, Phys. Rev.
A {\bf 38}, 2132 (1988).
\bibitem{Pindak89} J. W. Goodby {\it et. al.},  Nature
{\bf 337}, 449-452 (1989).
\bibitem{review}  H.G. Kitzerow, Chap. 10 of Chirality in
Liquid Crystals, Edited by H.G. Kitzerow and C. Bahr
(Springer-Verlag, New York, 2001).
\bibitem{deGennesSc}  P. de Gennes, Solid State Comm.
{\bf 14}, 997 (1973).
\bibitem{SC} A. A. Abrikosov, Sov. Phys. - JETP
{\bf 5}, 1174 (1957); P.G. de Gennes, {\it Superconductivity of
Metals and Alloys} (Benjamin, New York, 1966); Michael Tinkham,
{\it Introduction to Superconductivity} (McGraw Hill, New York,
1975).
\bibitem{RennLub-C}  T. C. Lubensky and S. R. Renn, Mol. Cryst. Liq.
Crys. {\bf 209}, 349-355 (1991).
\bibitem{RennC} S.R. Renn, Phys. Rev. A {\bf 45}, 953 (1992).
\bibitem{BTGBC} H.T. Nguyen et al. J. Phys. II (France) {\bf 2}
1889 (1992).
\bibitem{BandC}  L. Navailles, P. Barois, and H.T. Nguyen,
Phys. Rev. Lett {\bf 71},  545 (1993); L. Navailles, P. Barois,
and H.T. Nguyen, Phys. Rev. Lett. {\bf 72},  1300
  (1994).
\bibitem{squareTGBC} P.A. Pramod, Y. Hatwalne, N.V. Madhusudana,
Liq. Crystals {\bf 28} 525 (2001); N.A. Clark, private
communication.
\bibitem{Luk98} I. Luk'yanchuk, Physical Review E {\bf 57}, 574
(1998).
\bibitem{Dozov} I. Dozov, Phys. Rev.
Lett. {\bf 74}, 4245 (1995).
\bibitem{Chen-Lub} Jing-huei Chen, T. C. Lubensky, {\it Phys. Rev.
} A {\bf 14}, 1202 (1976).
\bibitem{MITsmc} L.J. Martinez-Miranda, A.R. Kortan, and R.J.
Birgeneau, Phys. Rev. Lett. {\bf 56}, 2264 (1986).
\bibitem{KL2} Arindam Kundagrami and T.C. Lubensky, unpublished.


\end{thebibliography}
\end{document}